\documentclass[12pt]{article}
\begin{document}

\title{Supersymmetry breaking and 4 dimensional string models}
\author{Michael Hewitt \\
Department of Computing,\\
Canterbury Christ Church University College,\\
North Holmes Road, Canterbury, CT1 1QU, U.K. \\ email:
mike.hewitt@canterbury.ac.uk,
 tel:+44(0)1227-767700.}
\date{11 February 2013}
\maketitle
\begin{abstract}

A family of superpotentials is constructed which may be relevant
to supersymmetry breaking in 4 dimensional (0,1) heterotic string
models. The scale of supersymmetry breaking, as well as the
coupling constant, would be stable.\

\


\end{abstract}

\section{Introduction}
\ \ \ In this paper we will study a family of superpotentials
which may be relevant to supersymmetry (susy) breaking in 4
dimensional (0,1) heterotic string models, and which generalise
those given in [1,2]. By 4 dimensional string models we have in
mind that the extra string degrees of freedom are subject to
twisted boundary conditions in such a way that no `breathing'
modes for extra dimensions appear in the level zero spectrum [3].
Stability of compactification is a feature of such $D=4$ string models, for which
decompactification cannot take place. The
gauge coupling, related to string topology through the genus
expansion, is dependent only on the dilaton multiplet in these models,
indicating a special role for this multiplet . The origin of susy breaking is often ascribed to
the gaugino condensation mechanism [4,5]. This relies on a
variation of the gauge coupling with a scalar field, which can
thus be identified as the dilaton in these models [6].  A vanishing, or almost vanishing, 4 dimensional
cosmological constant  seems possibly more natural in a model without
extra dimensions. For the present, we will consider the scalar
fields only for simplicity. A vanishing cosmological constant may
be viewed as a balance between local and global susy breaking.
This can be described geometrically as a requirement that the
field-space vector formed by all auxiliary fields be null. In
terms of the geometry of the chiral scalar fields, this is
equivalent to the superpotential $G$ having a critical slope of
$\sqrt{3}$ in field space. The polarization of positive and
negative energy auxiliary fields due to this local/global susy
breaking is much greater than the energy due to non-cancellation
between bosons and fermions from global susy breaking. A small
adjustment of the balance between gravitational and other
auxiliaries would cancel the residue to keep $\Lambda =0$. We will
assume that such a balancing mechanism applies. A small failure of
the balance mechanism could give a small $\Lambda$, as may be the
actual case in cosmology, but we will not attempt to understand
this here.

In $N=1$ susy, we may make a decomposition of the chiral fields
into the Goldstino $z$ and other fields $s_{i}$,where $z$,$s_{i}$
are orthogonal at the vacuum expectation values $z_{0},s_{0}$,
\begin{equation}
\frac{\partial^2 G}{\partial z \partial s^{*}_{i}}(z_{0},s_{0}) =
\frac{\partial G}{\partial s_{i}}(z_{0},s_{0}) = 0 \label{PDZ}
\end{equation}
We propose that the Goldstino scalar $z$ can be approximated by
the spin 0 part of the dilaton supermultiplet, which may be
regarded as scalar and pseudoscalar components of the graviton, as
we can form a massive gravitino multiplet, or extended graviton
multiplet by taking the full content of the spin $(1)_{L} \otimes
(1 \oplus 1/2)_{R}$ heterotic level zero fields. Our motivation is
that the dilaton multiplet in $D=4$ is naturally present, and has
a flat potential, or critical-slope superpotential (implementing
the local/global balance principle) in the absence of other
sources of susy breaking [7]. The pseudoscalar graviton mass is
zero, protected by a gauge invariance, and therefore the
pseudoscalar part of $z$ will be a pseudo-Goldstone boson with a
relatively long range. Mixings between the gravitational and other
scalars will be suppressed by factors of $1/m_{3/2}$, since the
contributions of different superfields to the Goldstone
supermultiplet would be in proportion to their auxiliary
components. CP violation may give a small scalar component to the
pseudo-Goldstone boson, but we will neglect this below.
 The situation in realistic models where there are also terms in $V$ from the gauge
auxiliary $D$ fields is more complicated, but providing that the
local/global balance principle applies, there will again be zero
cosmological constant with a small mixing $O(1/m_{3/2})$ between
$z$ and other superfields.
\section{Derivation of the Superpotentials}
\ \ \  Decomposing $z$ into scalar and pseudoscalar parts, we have
$z = x + \mathrm{i}y$ where
 \begin{equation}\label{y}
    dy  = *(dB + \omega_{GS}) + O(1/m_{3/2})
\end{equation}
($\omega_{GS}$ is the Green-Schwarz 3-form) so that under the
transformation $y \rightarrow y + \epsilon$, $\delta G = \epsilon
O(G/m_{3/2})$ or
\begin{equation}\label{z}
 \frac{\partial G(z)}{\partial y} = O(G/m_{3/2}),\  \frac{\partial G(z)}{\partial
 y}(z_{0},s_{0})=0
\end{equation}
i.e. $G(z) = G(x) + O(G/m_{3/2})$. For any function $h(z)$ which
depends only on $x$ we have
\begin{equation}\label{PD}
   \frac{\partial h(x)}{\partial z} = \frac{\partial h(x)}{\partial
   z^{*}} =  1/2 \frac{dh}{dx} \equiv \frac{h^{'}}{2}
\end{equation}
so that the potential for $x$ becomes (neglecting for now the
contribution from other fields)
\begin{equation}\label{V1}
 V(x) = \mathrm{e}^{G}(\frac{G^{'2}}{G^{''}}-3) =
 \frac{\mathrm{e}^{G}}{G^{''}}(G^{'2}-3G^{''}).
\end{equation}
Introducing the notation
\begin{equation}\label{g}
    G^{'} = g(x)
\end{equation}
we can write
\begin{equation}\label{f}
    g^{2} - 3g^{'} = f^{2}.
\end{equation}
For a vanishing cosmological constant, we require that $f$ should
have a zero. The simplest ansatz with this property is essentially
\begin{equation}\label{A}
f^{2} = \alpha (g-3)^{2}.
\end{equation}
Here we have used the linear transformation $x \rightarrow ax+b$
to arrange $x = 0, g = 3$ at $f = 0$ without loss of generality
(these values are chosen to simplify the formulas below). The case
$\alpha = 0$ gives the flat, no-scale solution

\begin{equation}\label{P}
G = -3\ln{(x)},\  \ V=0.
\end{equation}

We will now derive the solution corresponding to eq.(\ref{A}).
Rearranging eq.(\ref{f}) gives the first order differential
equation
\begin{equation}\label{x}
    \frac{dx}{dg} = \frac{1}{(1-\alpha)g^{2} + 6\alpha g -9\alpha}
\end{equation}
with solution
\begin{equation}\label{Sx}
    x = \tan^{-1}{(\frac{1-\alpha}{3\alpha}g + 1)} - \tan^{-1}{(\frac{1}{\alpha})}
\end{equation}
giving
\begin{equation}\label{Sg}
    g = \frac{3\alpha}{1-\alpha}(\tan{(x+c)-1)}
\end{equation}
where
\begin{equation}\label{c}
    c =  \tan^{-1}{(\frac{1}{\alpha})}.
\end{equation}
Now
\begin{equation}\label{G}
    G(x) = \int_{u = 0}^{x}g(u) du + G_{0}
\end{equation}
where the constant of integration $G_{0}$ gives the value of
$\ln{(m_{3/2}^{2})}$ at the potential minimum. We have
\begin{equation}\label{SG}
    G = G_{0}  + \frac{3\alpha}{1-
    \alpha}(\log \sec (x+c) - \log \sec (c)-x)
\end{equation}
and using
\begin{equation}\label{Gp}
G^{''} = g^{'} = \frac{3\alpha}{1-\alpha}\sec^{2}{(x+c)}
\end{equation}
gives us, on eliminating $c$ by using elementary identities,
\begin{equation}\label{SV2}
    V =3 \mathrm{e}^{G_{0}}\frac{(1+ \alpha^{2})}{1 - \alpha}[\mathrm{e}^{-x}(\cos{(x)} - \frac{1}{\alpha}\sin{(x)})]^{\frac{
    3\alpha}{\alpha -1}}\sin^{2}{(x)}.
\end{equation}
Here $x$ will be confined to the range where $V$ is finite i.e.
$\tan(x) < \alpha$ for $0 < \alpha < 1$. For the mass of the
scalar $x$ we have

\begin{equation}\label{ms}
m_{S}^{2} = \frac{V^{''}}{G^{''}}|_{x = 0}  =
\frac{\mathrm{e}^{G_{0}}}{G^{''}}
\frac{(f^{'})^{2}}{G^{''}}|_{x=0} = \alpha m_{3/2}^{2}
\end{equation}
giving the interpretation of the parameter $\alpha$ as
$m_{S}^{2}/m_{3/2}^{2}$. As argued above the mass of the
pseudoscalar $y$ will be
\begin{equation}\label{mp}
m_{P}^{2} = O(m_{3/2}^4).
\end{equation}
The solution eq.(\ref{SV2}) may be written implicity as a function
of the sigma model parameter

\begin{equation}\label{phi}
    \phi(x) = \int_{u=0}^{x}\sqrt{2G^{''}(u)}du =
    \sqrt{\frac{2\alpha}{3(1-\alpha)}}\ln{(\frac{\sec(x+c)+\tan(x+c)}{\sec(c)+ \tan(c)})}
\end{equation}
with standard kinetic term. Now consider the limit $\alpha
\rightarrow 1$ corresponding to the physically interesting case
$m_{S} = m_{3/2}$. Going back to eq.(\ref{f}) with $\alpha = 1$
gives

\begin{equation}\label{L}
   x = 1/2 \ln{(2g/3 - 1)}.
\end{equation}
Introducing an alternative parametrization by
\begin{equation}\label{w}
    w = \exp{(z)}, |w| = \exp{(x)}
\end{equation}
(so that $y$ and $y + 2\pi n$ are identified), which is natural
when we recall the origin of $x$ as the 4 dimensional dilaton,
gives the solution
\begin{equation}\label{SL}
    V = \frac{3 \mathrm{e}^{G_{0}}}{4}(|w|^{2} \exp{(|w|^{2}-1)})^{3/4}(|w| -
    |w|^{-1})^{2}.
\end{equation}
Here the value of $w$ is kept away from zero by the divergence in
$V(0)$. In this case the sigma model field is $\phi =
\sqrt{6}|w|$, so that if we write $\psi = |w|$ we can write the
Lagrangian terms for $\psi$ as

\begin{equation}\label{psi}
    L_{\psi} = 3\sqrt{-g}((\partial \psi)^{2} - \frac{\mathrm{e}^{G_{0}}}{4}
    (\psi^{2}\exp{(\psi^{2}-1}))^{3/4}(\psi - \psi^{-1})^{2}).
\end{equation}
(Here $g$ is the usual $D=4$ metric determinant.) An interesting
feature of this model is that the scalar and pseudoscalar fields
will be described by a sigma model with the flat Kahler metric

\begin{equation}\label{sig}
    d \sigma^{2} = 6dwdw^{*} = 6(d\psi^{2} + \psi^{2}d\theta^{2}).
\end{equation}

 We will now consider the origin of the integration constant $G_{0}$, which determines $m_{3/2}$.
 Suppose that, near $(z_{0},s_{0})$, $G$ takes the form
\begin{equation}\label{K}
    G(z,s) = G_{g}(z)+ G_{s}(s) + H(z,s) = G_{g}(z)+ K(s) +
    \ln(|W(s)|^{2}) + H(z,s)
\end{equation}
solving eq.(\ref{PDZ}), where $G_{g}(z)$ takes the form discussed
above and for the remaining chiral fields $s$, $K(s)$ is the
Kahler potential (generating the field kinetic terms) and $W(s)$
represents the chiral scalar superfield interactions. $H(z,s)$
represents mixing terms of higher order in $1/m_{3/2}$ relative to
$G(z,s)$, with $H(z_{0},s_{0})=0$. We expect $K_{0} \sim 0$,
giving
\begin{equation}\label{W}
    \exp{(G_{0})} \sim |W_{0}|^{2}
\end{equation}
where $W_{0}$ is the minimum value of $W(s)$ - since the vacuum
represents the minimum of $G$ w.r.t. $s$, it also represents the
minimum of $W(s)$. Now the leading renormalizable contribution
gives $W \sim g_{4}X_{1}X_{2}X_{3}$ where $X_{i}$ may acquire
expectation values by a Coleman-Weinberg mechanism well below the
string tension scale. For $m_{3/2}\sim$ 1TeV, the $X_{i0}$ may
come from GUT symmetry breaking, a neutrino mass see-saw mechanism
or some other hidden sector. String diagrams with zero background
field give $V(z) = 0$. While this may represent the $\alpha
\rightarrow 0$ limit of the family of models represented by
eq.(\ref{A}), it may alternatively represent the limit
$\exp{(G_{0})} \rightarrow 0$, in particular with $\alpha = 1$,
allowing stable supersymmetry breaking in 4 dimensional string
models. For $\alpha \leq 1$, $m_{3/2}$ is prevented from running
away to zero, so that it is natural in this scenario for susy
breaking to survive inflation.

\

\textbf{References}

[1] J.Polonyi, Budapest Preprint KFKI-1997-93.

[2] E.Cremmer, S.Ferrara, C.Kounnas, and D.V.Nanopoulos, {\em
Phys. Lett B\/}{\bf 133} (1983) 61.

[3] H.Kawai, D.C.Lewellen and S.H.H.Tye, {\em Nucl. Phys. B\/}{\bf
288} (1987) 1.

[4] S.Ferrara, L.Girardello and H.P.Nilles, {\em Phys. Lett
B\/}{\bf 125} (1983) 457.

[5] M.Dine, R.Rohm, N.Seiberg and E.Witten, {\em Phys. Lett
B\/}{\bf 156} (1985) 55.

[6] E.S.Fradkin, A.A.Tseytlin, {\em Phys. Lett B\/}{\bf 160}
(1985) 69.

[7] M.Dine and N.Seiberg, {\em Phys. Rev. Lett\/}{\bf 57} (1986)
2625.

\end{document}